\documentclass[a4paper,12pt]{article}
\usepackage{amsmath}
\usepackage{amssymb}
\usepackage{latexsym}
\usepackage{mathrsfs}

\def \d {\mathrm{d}}

\begin{document}

\title{{\bf {\Large General Notion of Curvature \\ in Catastrophe Theory
Terms}}}
\author{
Petko A. Nikolov {\normalsize $^{\!\!\!{\,}^{a)}}$}\footnote{
pnikolov@phys.uni-sofia.bg }\quad Lora Nikolova {\normalsize
$^{\!\!\!{\,}^{b)}}$}\footnote{ lora@inrne.bas.bg }\quad Gergana
R. Ruseva {\normalsize $^{\!\!\!{\,}^{b)}}$}\footnote{
gergana@inrne.bas.bg }
\\ \\
{\normalsize \small \( \begin{array}{l} {\!\!\!\!\!}{\,}^{a)} \,
\mathrm{ Department \ of \ Theoretical \ Physics, \ Sofia \
University } \\ \, \mathrm{ James \ Bourchier \ 5, \ BG \!\! -
\!\! 1164 \ Sofia, \ Bulgaria }
\\ \\ {\!\!\!\!\!}{\,}^{b)} \,
\mathrm{ Institute \ for \ Nuclear \ Research \ and \ Nuclear \
Energy, } \\ \, \mathrm{ Tsarigradsko \ Chaussee \ 72, \ BG \!\! -
\!\! 1784 \ Sofia, \ Bulgaria }
\end{array} \) }}

\maketitle

\begin{abstract}

We introduce a new notion of a curvature of a superconnection,
different from the one obtained by a purely algebraic analogy with
the curvature of a linear connection. The naturalness of this new
notion of a curvature of a superconnection comes from the study of
singularities of smooth sections of vector bundles (Catastrophe
Theory). We demonstrate that the classical examples of
obstructions to a local equivalence: exterior differential for
2-forms, Riemannian tensor, Weil tensor, curvature of a linear
connection and Nijenhuis tensor can be treated in terms of one
general approach. This approach, applied to the superconnection
leads to a new notion of a curvature (proposed in this paper) of a
superconnection .
\end{abstract}

\section{A Brief Review of the Notion of a Superconnection}\label{sec:1}

The notions of a superconnection and of the corresponding
supercurvature were introduced by Daniel Quillen in 1985
\cite{Quillen}. In this section we give a brief review of the
matter and introduce the basic notations.

By $\xi = (E, p, M)$ we denote a vector bundle over the manifold
$M$ $(\dim M=m$, $\dim (\xi)=n)$, by $\xi^*$ - the dual bundle and
by $C^\infty(\xi)$ - the space of the vector fields, i.e. the
space of the smooth sections of the bundle $\xi$. $\Omega^k(M)=
C^\infty(\Lambda^kT^*(M))$ is the the space of the differential
$k$-forms on the manifold $M$ and
\[\Omega(M) = \mathop{\oplus}_{k=0}^{m} \Omega ^k(M) = C^\infty
(\mathop{\oplus}_{k=0}^{m} \Lambda^kT^*(M))\] is the space of the
nonhomogeneous forms on $M$. $ \Omega^k(\xi) = C^\infty
(\Lambda^kT^*(M) \otimes \xi)$ is the space of the differential
$k$-forms with values in $\xi$ and $\Omega(\xi) =
\mathop{\oplus}_{k=0}^{m}\Omega^k(\xi)$ is the space of the
nonhomogeneous forms with values in $\xi$. By a linear connection
$\nabla$ on $\xi$ we understand a linear differential operator
$\nabla: \Omega^0(\xi) \longrightarrow \Omega^1(\xi)$ with the
property $\nabla(f.\psi) = \d f \otimes \psi + f\nabla(\psi)$,
$\psi \in \Omega^0(\xi)$, $f \in \Omega^0(M)$. The space of the
linear connections is an affine space with a linear group
$\Omega^1(\xi^* \otimes \xi)$. If we choose an arbitrary linear
connection $\nabla_0$ as "an origin" then for every linear
connection $\nabla$ on $\xi$ we have
\begin{equation}\label{1}
\nabla = \nabla_0 +A \, , \, \, A \in \Omega^1(\xi^* \otimes \xi).
\end{equation}

The connection $\nabla$ generates a covariant differential
$\d^\nabla : \Omega^k(\xi) \longrightarrow \Omega^{k+1}(\xi)$
defined by the property $\d^\nabla(\alpha \otimes \psi) = \d
\alpha \otimes \psi + (-1)^k \d \wedge \nabla (\psi)$, $\alpha \in
\Omega^k(M)$, $k= 1,2,...,m$ and $\psi \in \Omega^0(\xi)$. Its
square $\d^\nabla \circ \d^\nabla : \Omega^k(\xi) \longrightarrow
\Omega^{k+2} (\xi)$ is an $\Omega^0(M)$-linear operator and as a
consequence $F^\nabla = \d^\nabla \circ \d^\nabla: \Omega^0(\xi)
\longrightarrow \Omega^2(\xi)$ is a differential operator of zero
order i.e. $F^\nabla$ is a tensor, $F^\nabla \in \Omega^2(\xi^*
\otimes \xi)$. Let $\{x^\mu\}$ denotes (local) coordinates on $M$,
$\{e_a\}$ - a (local) basis in $\xi$ and $\{e^a\}$ - the
corresponding dual basis in $\xi^*$. In local coordinates
\begin{equation}\label{2}
(\nabla(\psi))_\mu^a = \partial_\mu \psi^a + A_{\mu b}^a \psi^b
\end{equation}
and
\begin{equation}\label{3}
F_{\mu\nu \, b}^{\nabla \, a} = (\partial_\mu A_\nu -
\partial_\nu A_\mu + [A_\mu, A_\nu])^a_b\,.
\end{equation}

The upper notions have an algebraic analogy in the case of a
$\mathbb{Z}_2$ graded super vector bundle \cite{Quillen}. Let $\xi
= \xi_+ \oplus \xi_-$ be a $\mathbb{Z}_2$ graded vector bundle
over the manifold $M$. Then the dual bundle $\xi^* = \xi^*_+
\oplus \xi_-^*$ is also $\mathbb{Z}_2$ graded in a natural way.
The induced $\mathbb{Z}_2$ grading in $\Omega(\xi)$ is given by
\[
\Omega(\xi) = \Omega(\xi) _+ \oplus \Omega(\xi)_- \, , \] where
\[ \Omega(\xi)_+ = \mathop{\oplus}_{k}(\Omega^{2k}(\xi_+)
\oplus \Omega^{2k+1}(\xi_-))\] and
\[
\Omega(\xi)_- = \mathop{\oplus}_{k}(\Omega^{2k+1}(\xi_+) \oplus
\Omega^{2k}(\xi_-))\, .\] Let $\nabla$ be a linear connection on
the vector bundle $\xi$ compatible with the $\mathbb{Z}_2$ grading
of $\xi$,
\[
\nabla: \Omega^0(\xi_{\pm}) \longrightarrow \Omega^1(\xi_\pm) \, ,
\]
i.e. $\nabla$ is an odd linear operator.

Let $\chi \in \Omega^0(\xi^* \otimes \xi)$ be an odd tensor field,
i.e. $\chi(x) : \xi_{x\pm} \longrightarrow \xi_{x\mp}$ (or
$\chi(x) \in (\xi^* \otimes \xi)_{x-}).$ By definition (proposed
by Quillen) a superconnection $\nabla_s$ is the odd linear
operator
\[
\nabla_s = \nabla + \chi: \Omega^0(\xi)_\pm \longrightarrow
\Omega^{0, 1}(\xi)_\mp \quad ,
\]
where $ \Omega^{0, 1}(\xi) = \Omega^{0}(\xi) \oplus
\Omega^{1}(\xi) \, .$ It is easy to see that the superconnection
$\nabla_s$ has the property
\[
\nabla_s(f.\psi) = df \otimes \psi + f\nabla_s(\psi) \, , f \in
\Omega^0(M) \, , \psi \in \Omega^1(\xi) \, .
\]
In more details, if we write $\psi = \left[
\begin{array}{c}\psi_+ \\ \psi_-
\end{array} \right]\, ,$ where $\psi_+$ and $\psi_-$ are the even
and the odd parts of the super vector field $\psi$ respectively,
then
\[
\nabla_s \left[ \begin{array}{c} \psi_+ \\ \psi_- \end{array}
\right] = \left[ \begin{array}{cc} \nabla & \chi_{+-} \\ \chi_{-+}
& \nabla \end{array} \right] \left[ \begin{array}{c} \psi_+ \\
\psi_- \end{array} \right] = \left[ \begin{array}{c} \nabla \psi_+
+ \chi_{+-} \psi_-
\\
\nabla\psi_- + \chi_{-+} \psi_+ \end{array} \right] \, .
\]
The space of the superconnections is again an affine space and if
we choose an arbitrary superconnection $\nabla_{s0}$ as an origin,
then for any other superconnection $\nabla_s$ we have
\[
\nabla_s = \nabla_{s0}+ A_s \, , \quad A_s\in \Omega^{0,1}(\xi^*
\otimes \xi)_- \,.
\]
The covariant differential
\[
\d^{\nabla_s} : \Omega ^k(\xi) \longrightarrow \Omega^{k, k+1}
(\xi)
\]
can be defined by a purely algebraic analogy with the "classical"
case,
\[
\d^{\nabla_s}(\alpha \otimes \psi) = \d\alpha \otimes \psi +
(-1)^k \alpha \wedge \nabla_s (\psi) \, ; \, \alpha \in
\Omega^k(M)\, , \, \psi \in \Omega^0 (\xi) \,.
\]
It is easy to see that again $\d^{\nabla_s} \circ \d^{\nabla_s}:
\Omega^k(\xi) \longrightarrow \Omega^{k, k+1, k+2}(\xi)$ is an
$\Omega^0(M)$ linear operator. $F^{\nabla_s} = \d^{\nabla_s} \circ
\d^{\nabla_s}: \Omega^0(\xi) \longrightarrow \Omega^{0, 1, 2}
(\xi)$ is an even differential operator of order zero, i.e.
$F^{\nabla_s}$ is a tensor, $F^{\nabla_s} \in \Omega^{0, 1, 2}
(\xi^* \otimes \xi)\, , $
\[
 F^{\nabla_s}  = \chi^2 +
\d^\nabla(\chi) + F^\nabla \, .
\]
The nonhomogeneous tensor $F^\nabla$ defined by an algebraic
analogy with the "classical" case is, by definition, the
supercurvature of the superconnection $\nabla_s \, .$

In the classical case the connection $\nabla$ is connected with
the parallel transport and the curvature is an obstruction to the
flatness of the parallel transport. In other words the curvature
of a connection is an obstruction to its local equivalence with
the flat connection. In the "super" case there are no natural
notions of a parallel transport and of a flat superconnection. But
we can look for an obstruction to the local equivalence of two
superconnections. To motivate the idea we consider in the next
section some classical examples of obstructions to a local
equivalence of sections of some bundles as particular examples of
a general scheme.

\section{A List of Obstructions to a Local Equivalence}\label{sec:2}

\subsection{An Obstruction for an Arbitrary Nondegenerate Differential 2-form to be Diffeomorphic to the Canonical Symplectic Form}\label{sec:2.1}

Let $M$ be a smooth even-dimensional manifold, $\dim M = m = 2l$
and let $\omega_0 \in \Omega^2(M)$. In local coordinates $
\omega_0 = \mathop{\sum}_{\mu=1}^{l}\d x^\mu \wedge \d x^{l+\mu}\,
.$ Every diffeomorphism $\varphi \in Diff (M)$ has a natural
action on $\Omega^2(M)\, .$ For every $\omega \in \Omega^2(M)$ we
have
\[
\varphi^*(\omega)_{\mu \nu} (x) = \frac{\partial
\varphi^\alpha}{\partial x^\mu}(x) \frac{\partial
\varphi^\beta}{\partial x^\nu} (x) \omega_{\alpha \beta}(\varphi
(x)) \, .
\]
Does there exist (at least locally) $\varphi \in Diff(M)$ such
that $\varphi^*(\omega) = \omega_0$ in the case $\det
\{\omega_{\mu \nu} (x) \} \neq 0 \,?$ An obstruction to this is
the condition $\d \omega \neq 0 $. If $\d \omega = 0$ then there
exists (at least locally) $\varphi \in Diff (M)$ such that
$\varphi^*(\omega) = \omega_0 \, .$ The 3-form $\d \omega$ is in
some sense the curvature of the 2-form $\omega$.

\subsection{An Obstruction for an Euclidean Metric to be in a Canonical Form}\label{sec:2.2}

Let $M$ be a smooth manifold, $g_0 = \mathop{\sum}_{\mu = 1}^{m}
\d x^\mu \otimes \d x^\mu$ - an Euclidean metric on $M$ in a
canonical form and $g$ - an arbitrary Euclidean metric on $M$.
Does there exist (at least locally)  $\varphi \in Diff (M)$ such
that
\[
\varphi^*(g)_{\mu \nu}(x) = \frac{\partial
\varphi^\alpha}{\partial x^\mu}(x) \frac{\partial
\varphi^\beta}{\partial x^\nu} (x) g_{\alpha \beta}(\varphi(x)) =
g_{0 \mu \nu}(x) = \delta_{\mu \nu} \, ?
\]
The Riemannian tensor $R(g) \neq 0 $ is an obstruction to this. If
$R(g) = 0$ then there exists (at least locally) a $\varphi \in
Diff(M)$ such that $\varphi^*(g) = g_0$. The Riemannian tensor
$R(g)$ is the curvature of the metric $g$.

\subsection{An Obstruction for an Euclidean Metric to be Conformaly Equivalent to the Canonical Metric}\label{sec:2.3}

Let $g_0$ and $g$ be the same as in Subsection 2.2. Does there
exist (at least locally) $\varphi \in Diff (M)$ such that
$\varphi^*(g)(x) = f(x) g_0(x)\, ,$ $f(x) \neq 0\, ?$ The Weil
tensor
\[
W(g) = R(g) - \frac{1}{m-2} (Ric(g) - \frac{1}{2(m-2)} r(g).g)
\wedge g \neq 0\,,
\]
where $Ric(g)$ is the Ricci tensor, $r(g)$ - the scalar curvature,
is an obstruction to this \cite{Besse}. If $W(g) = 0$ then there
exists (at least locally) $\varphi \in Diff(M)$ such that
$\varphi^*(g)(x) = f(x) g_0(x)$; $f(x) \neq 0$. The Weil tensor
$W(g)$ is the conformal curvature of the metric $g$.

\subsection{An Obstruction for a Given Connection to be Gauge
Equivalent to the Flat One}\label{sec:2.4}

Let $\xi = (E, p, M)$ be a vector bundle over the manifold $M$ and
$\nabla$ - a linear connection on $\xi \,$ (see (\ref{1})). In
local coordinates we have
\[
\nabla_\mu = \partial_\mu + A_\mu \,.
\]
Let $\varphi \in Aut^V (\xi)$ be a vertical automorphism of $\xi$.
The automorphism $\varphi$ has a natural action on the space of
the linear connections. In local coordinates \[
 \varphi^* (\nabla)_\mu = \partial_\mu + \varphi^*(A)_\mu
\, , \, \varphi^*(A)_\mu(x) = \varphi^{-1}(x). A_\mu(x)
.\varphi(x) + \varphi^{-1}(x).\partial_\mu\varphi(x) \, . \]

Can we find $\varphi \in Aut^V(\xi)$ satisfying the condition
$\varphi^*(A)_\mu = 0$? The Yang-Mills curvature tensor $\{F_{\mu
\nu}\} \neq 0$ is an obstruction to this. If $F_{\mu \nu} =
\partial_\mu
 A_\nu - \partial_\nu A_\mu +[A_\mu, A_\nu] = 0$ we can find (at
least locally) an automorphism $\varphi \in Aut^V(\xi)$ such that
\[
\varphi^*(A)_\mu(x) = \varphi^{-1}(x). A_\mu(x). \varphi(x) +
\varphi^{-1}(x).\partial_\mu \varphi(x) = 0 \, .
\]
The Yang-Mills tensor $F^\nabla$ is the curvature of the linear
connection $\nabla$.

\subsection{An Obstruction for a Given Almost Complex Structure to be
Diffeomorphic to the Canonical One}\label{sec:2.5}

Let $M$ be an even dimensional smooth manifold, $\dim M = m = 2l$
and let
\[
J_0 = \mathop{\sum}_{i=1}^{l}\left(\d x^i \otimes
\frac{\partial}{\partial x^{2+i}} - \d x^{l+i} \otimes
\frac{\partial}{\partial x^i} \right)
\]
 be an almost complex
structure on $M$ in a canonical form in local coordinates. Let $J$
be an arbitrary almost complex structure on $M$ \cite{Kolar}, $J
\in C^\infty (T^*(M) \otimes T(M))$, and $J^2(x) = -\mathbf{1}$.
In the same coordinates,
\[
J(x) = J^\mu_\nu(x) \d x^\nu \otimes \frac{\partial}{\partial
x^\mu }\, .
\]
Every diffeomorphism $\varphi \in Diff(M)$ has a natural action on
$C^\infty(T^*(M) \otimes T(M))$. For every $J \in C^\infty(T^*(M)
\otimes T(M))$ we have
\[
\varphi^*(J)_\nu^\mu(x) = \frac{\partial \varphi^{-1\mu}}{\partial
x^\alpha}(\varphi(x))\frac{\partial \varphi^{\beta}}{\partial
x^\nu}(x)J_\beta^\alpha(\varphi(x))\, .
\]
Does there exist $\varphi \in Diff(M)$ such that $\varphi^*(J) =
J_0$? The nonvanishing Nijenhuis tensor,
\[
N(J)_{\mu \nu}^\rho = J_\mu^\alpha \partial_\alpha J_\nu^\rho -
J_\nu^\alpha \partial_\alpha J_\mu^\rho - J_\alpha^\rho
\partial_\mu J_\nu^\alpha + J_\alpha^\rho
\partial_\nu J_\mu^\alpha
\]
is an obstruction to this. If $N(J) = 0$ then there exists (at
least locally) $\varphi \in Diff(M)$ such that $\varphi^*(J) =
J_0$ (see Newlander - Nirenberg theorem \cite{Newlander}). The
Nijenhuis tensor can be written as a vector-valued 2-form
\[
N(J)(X,Y) = [JX, JY] - J[X, JY] - J[JX, Y] - [X, Y] \, .
\]
The Nijenhuis tensor $N(J)$ is in some sense the curvature of the
almost complex structure $J$.

All the obstructions to the local equivalence of sections in the
corresponding fibre bundles with respect to the action of the
"functional" group  we mentioned in the upper cases can be viewed
as "curvatures" in the broad sense of the word. We can consider
them as particular examples of a general notion of a curvature
(see Section 3) considered as an universal type of obstructions to
the local equivalence in terms of jet bundles, jet lifting of the
group action and the algebraization of the differential operators.
This scheme applied to the case of a superconnection leads to a
notion of a curvature of a superconnection which is different from
the one, obtained by a purely algebraic analogy \cite{Quillen}. We
consider the definition given in Section 4 to be the adequate one.

\section{General Notion of a Curvature}\label{sec:3}

For a vector bundle $\xi = (E, p, M)$ the adopted coordinates
$(x^\mu, u^a) $ on $\xi $ induce the corresponding coordinates
$(x^\mu, u^a, u_\nu^a, u_{\nu_1 \nu_2}, ..., u^a_{\nu_1, ...,
\nu_k})$, $1 \leq \nu_1 \leq \nu_2 \leq ... \leq \nu_j \leq m $, $
j=2, 3, ..., k $ in the k-th jet bundle $j^k(\xi)$ of $\xi$,
$j^k(\xi) = (E^k, p^k, M)$ for every $k= 1,2,3,...$ (see
\cite{Palais}, \cite{Saunders}). Every element of the fibre
$j^k(\xi)_x$ of $j^k(\xi)$ equals to $j^k(\psi(x))$ - the value at
the point $x$ of the $k$-jet of some $\psi \in C^\infty(\xi)$, in
coordinates $u^a(j^k(\psi(x)) = \psi^a(x)$, $u^a_{\nu_1 ...
\nu_j}(j^k(\psi(x))) =
\partial_{\nu_1}...
\partial_{\nu_j} \psi^a(x)$, where $\psi^a$ are the components of
the field $\psi$. The mapping $C^\infty(\xi) \longrightarrow
C^\infty(j^k(\xi))$, given by $\psi(x) \longmapsto (\psi^a(x),
\partial_\mu \psi^a(x), ...,
\partial_{\mu_1}...\partial_{\mu_k} \psi^a(x))$ is the so called
jet lifting of the section $\psi$ and plays the role of an
universal differential operator of order $k$ (see \cite{Palais},
\cite{Saunders}).

Let $H$ be a "functional" group and $\textit{F}_h:E
\longrightarrow E$, $h \in H$ be a fibre preserving action of the
group $H$ on the bundle $\xi = (E, p, M)$. For sake of simplicity
we will have in mind only the group $Aut^V(\xi)$ of vertical
automorphisms of $\xi$, or $Diff(M)$ - the group of
diffeomorphisms of $M$. In the coordinates $(x^\mu, u^a)$ the
action $\textit{F}$ of the group $H$ reads
\[
\textit{F}_h(x^\mu, u^a) = (\varphi_h^\mu(x), F_h^a(x, u))\, ,
\]
where $h \in H$, $\varphi_h: M \longrightarrow M$ is an action of
the group $H$ on the base $M$ of the bundle and
$F_{h_1}(\varphi_{h_2}(x), F_{h_2}(x, u)) = F_{h_1h_2}(x, u) \,.$

The group $H$ has a natural action $\tilde{\textit{F}}$ on
$C^\infty$
\begin{equation} \label{10}
\tilde{\textit{F}}_h(\psi)(x) = F_h(\psi(\varphi_h^{-1}(x)))\, ,
\end{equation}
where $\psi \in C^\infty(\xi)$, $h \in H$, $x \in M$. The action
(\ref{10}) of the group $H$ on  $C^\infty (\xi)$ induces an action
of $H$ on the jet-bundle $j^k(\xi)$ for every $k = 1, 2, 3, ...$.

Let $\psi_1$ and $\psi_2$ be two sections of the bundle $\xi$. The
problem we deal with is the (local) equivalence of $\psi_1$ and
$\psi_2$ with respect to the group $H$, i.e. the existence of an
element $h \in H$ such that
\begin{equation}\label{11}
\tilde{\textit{F}}_h(\psi_2) = \psi_1 \,.
\end{equation}
In other words for given $\psi_1$ and $\psi_2$, the expression
(\ref{11}) is an equation for $h\in H$. If the equation (\ref{11})
is satisfied then its $k$-jet lifting
\begin{equation}\label{12}
\tilde{\textit{F}}_h(j^k (\psi_2)) = j^k(\psi_1) \,.
\end{equation}
is also satisfied for every $k$. Inversely, if we can proof that
for some $k$ the equation (\ref{12}) has no solution for $h \in H$
then obviously this is an obstruction to the solvability of
(\ref{11}), i.e. to the local equivalence of $\psi_1$ and
$\psi_2$. The equation (\ref{12}) involves the derivatives of
$\psi_1$ and $\psi_2$ up to the order $k$. Therefore it is easier
to deal with (\ref{12}).

We begin with the study of the condition imposed by the equation
(\ref{12}) on the $k$-jets of $\psi_1$ and $\psi_2$ at a point
$x_0 \in M$.

Let
\[
G_{x_0}:= \{h \in H \mid \varphi_h(x_0) = x_0 \}
\]
be the stationary group of the point $x_0$. The group $G_{x_0}$
has a natural action on $j^k(\xi)_{x_0}$ for every $k$. Let us
consider first the case $k=0$. The space $j^0(\xi)_{x_0}$ is
simply the fibre $\xi_{x_0}$. For $h \in G_{x_0}$ the equation
(\ref{12}) leads to
\begin{equation}\label{13}
\tilde{\textit{F}}_h(\psi_2(x_0)) = \psi_1(x_0) \,.
\end{equation}
Insolvability of (\ref{13}) with respect to $h$ means that
$\psi_1(x_0)$ and $\psi_2(x_0)$ belong to different orbits of
$G_{x_0}$ in $\xi_{x_0}$. If the fibre $\xi_{x_0}$ is a
homogeneous space for the group $G_{x_0}$ there is no obstruction
to the solvability of (\ref{12}) arising from the equation
(\ref{13}). Then looking for an obstruction we go to $k=1$:
\begin{equation}\label{14}
\tilde{\textit{F}}_h(j^1 (\psi_2)_{x_0}) = j^1(\psi_1)_{x_0} \,.
\end{equation}
If the fibre $j^1(\xi)_{x_0}$ is again a homogeneous space for the
group $G_{x_0}$ there is no an obstruction to the solvability of
equation (\ref{12}). We proceed to the first $k$ for which the
fibre $j^k(\xi)_{x_0}$ is not a homogeneous space for the group
$G_{x_0}$. Let $\tilde{\pi}_{x_0}$ be the canonical projection to
the factor space $j^k(\xi)_{x_0}/G_{x_0}$
\begin{equation}\label{15}
\tilde{\pi}_{x_0} : j^k (\xi)_{x_0} \longrightarrow j^k
(\xi)_{x_0}/_{G_{x_0}}\,.
\end{equation}
If
\begin{equation}\label{16}
\tilde{\pi}_{x_0} ( j^k (\psi_2)_{x_0}) \neq \tilde{\pi}_{x_0} j^k
(\psi_1)_{x_0})
\end{equation}
then the $k$-jets $j^k (\psi_2)_{x_0}$ and $j^k (\psi_1)_{x_0}$
belong to different orbits of the group $G_{x_0}$ and the equation
\begin{equation}\label{17}
\tilde{\textit{F}}_h(j^k (\psi_2))_{x_0} = j^k(\psi_1)_{x_0}
\end{equation}
has no solution with respect to $h\in G_{x_0}$. The condition
(\ref{17}) is an obstruction to the local equivalence of the
sections $\psi_2$ and $\psi_1$ in a neighborhood of the point
$x_0$. These considerations are valid for every point $x \in M$
and we obtain a fibre preserving map (over the identity)
\[
\tilde{\pi}: j^k(\xi) \longrightarrow j^k(\xi)/H \, ,
\]
where the fibre of $j^k(\xi)/H$ at a point $x \in M$ is the factor
space $j^k(\xi)_x/G_x$. The fibre preserving map $\tilde{\pi}$ is
the symbol of the differential operator
\[
\pi : C^\infty(\xi) \longrightarrow C^\infty (j^k(\xi)/H)\,,
\]
i.e. $\pi(\psi) = \tilde{\pi}(j^k(\psi))$. The condition
$\pi(\psi_2) \neq \pi(\psi_1)$ is an obstruction to the (local)
equivalence. Therefore we consider the differential operator $\pi$
as a curvature - the general notion of curvature that we propose
in this paper ($\pi(\psi)$ is the curvature of the section
$\psi$).

One can show that the classical results listed in Section 2 are
explicit examples of the differential operator $\pi$. Namely:
\begin{enumerate}
    \item[\textbf{2.1.}] The sections we consider are the two forms $\omega$ on $M$,
    the functional group is the group $Diff(M)$, $k=1$ and
$\pi(\omega)
    = \d \omega$.
    \item[\textbf{2.2.}] The sections we consider are the metrics $g$ on $M$, the
    functional group is the group $Diff(M)$, $k=2$ and $\pi(g) =
    R(g)$ - the Riemannian curvature tensor.
    \item[\textbf{2.3.}] The sections we consider are the conformaly equivalent
    classes $[g]$ of metrics $g$, the functional group is
    $Diff(M)$, $k=2$ and $\pi([g]) = W(g)$ - the Weil conformal
    tensor.
    \item[\textbf{2.4.}] The sections we consider are the linear connections
    $\nabla$ on a vector bundle $\xi$, $k=1$ and $\pi(\nabla) =
    F^\nabla$ - the Yang-Mills curvature tensor.
    \item[\textbf{2.5.}] The sections we consider are the almost complex
    structures $J$ on $M$, the functional group is
    $Diff(M)$, $k=1$ and $\pi(J) \sim N(J)$ - the Nijenhuis
    tensor.
\end{enumerate}

In the next section we will consider the cases \textbf{2.4} and
\textbf{2.5}.

\textbf{Remark:} Here we consider two kinds of groups.

1. The group $H$ is the group of vertical automorphisms of a
vector bundle $\xi$, $H=Aut^V(\xi)$. In this case the stationary
group of a point $x_0$, $G_{x_0} = H$.

2. The group $H$ is the group of the diffeomorphisms of a manifold
$M$, $H= Diff(M)$. The bundle $\xi$ is some tensor power of $T(M)$
and of $T^*(M)$ and the action of $H$ is its tangent lifting. In
this case it is enough to consider only its stationary group $G_x$
at each point $x$. This matter will be discussed in a following
paper.

\section{Examples}\label{sec:4}

\textbf{Example} \textbf{2.4} - Linear connections on a vector
bundle.

We will describe the curvature of a linear connection on a vector
bundle as an obstruction to the local equivalence of linear
connections in terms of the general scheme given in Section 3.

For every linear connection $\nabla$ on a vector bundle $\xi$ we
have $\nabla_\mu = \partial_\mu + A_\mu$, where $\nabla_{0\mu} =
\partial_\mu$ is the "origin" and $\{A_\mu\} \equiv A \in
\Omega^1(\xi^* \otimes \xi)$. The elements of $\Omega^1(\xi^*
\otimes \xi)$ are in one-to-one correspondence with the linear
connections on the bundle $\xi$ and we recognize $\Omega^1(\xi^*
\otimes \xi)$ as the space of sections of $T^*(M) \otimes \xi^*
\otimes \xi$, where the "functional" group $Aut^V(\xi)$ acts. To
accomplish the procedure described in section 3, we choose
coordinates $\{x^\mu, u^a\}$ to be centered at a point $x_0 \in
M$, $x^\mu(x_0) = 0$. The 1-jet of $A_\mu$ at the point $0$ reads
\begin{equation}\label{25}
j^1(A_\mu)_0(x) = \textbf{A}_\mu + \textbf{A}_{\mu \alpha}x^\alpha
\, ; \, \textbf{A}_\mu = A_\mu(0) \, , \, \textbf{A}_{\mu \alpha}
= \partial_\alpha A_\mu (0) \, ,
\end{equation}
where $\mu, \alpha = 1, 2, ..., m$. The set of all pairs of
arbitrary matrices $(\{\textbf{A}_\mu\}, \{\textbf{A}_{\mu
\alpha}\})$ describes $j^1(T^*(M) \otimes \xi^* \otimes \xi)_0$.
Let $\varphi \in Aut^V(\xi)$ be a vertical automorphism, $\varphi
(x^\mu) \in GL(n, \mathbb{R})$. The 2-jet of $\varphi$ at the
point $0$ reads
\begin{equation}\label{29}
j^2(\varphi)_0(x) = \textbf{B}_0 +\textbf{B}_\alpha x^\alpha +
\frac{1}{2} \textbf{B}_{\alpha \beta} x^\alpha x^\beta \, ;
\textbf{B}_0 = \varphi (0) \, , \, \textbf{B}_{\alpha \beta} =
\partial_\alpha \partial_\beta \varphi (0) \, ,
\end{equation}
where $\alpha, \beta = 1, 2, ..., m$. The set of all triples of
arbitrary matrices $\{\textbf{B}_0$ with $\det (\textbf{B}_0) \neq
0$, $\textbf{B}_\alpha $, $\textbf{B}_{\alpha \beta}$ symmetric
with respect to $\alpha , \beta \}$ describes the space of 2-jets
of the vertical automorphisms at the point $0$. The matrix
$\textbf{B}_0 = \varphi (0)$ is non-degenerate and can be
considered as a common multiplier of the entire 2-jet and it does
not play an essential role in our considerations. It is enough to
consider only automorphisms $\varphi$ with 2-jets of the form
\begin{equation}\label{30}
j^2 (\varphi)_0(x) = \textbf{1} + \textbf{B}_\alpha x^\alpha +
\frac{1}{2} \textbf{B}_{\alpha \beta} x^\alpha x^\beta \, .
\end{equation}
The action of $\varphi$ by its 2-jet (\ref{30}) on $j^1(T^*(M)
\otimes \xi^* \otimes \xi)_0$ is given by
\[
j^1(A_\mu)_0 \mapsto j^1(\varphi^*(A)_\mu)_0 = j^1(\varphi^{-1})_0
. j^1(A_\mu)_0 . J^1(\varphi)_0 + j^1 (\varphi^{-1})_0.
j^1(\partial_\mu \varphi)_0 \, .
\]

\begin{equation}\label{35}
\left|
\begin{array}{l} \textbf{A}_\mu \mapsto \textbf{A}_\mu +
\textbf{B}_\mu \\
\textbf{A}_{\mu \alpha} \mapsto \textbf {A}_{\mu \alpha} +
\textbf{A}_\mu . \textbf{B}_\alpha  - \textbf{B}_\alpha .
\textbf{A}_\mu - \textbf{B}_\alpha . \textbf{B}_\mu +
\textbf{B}_{\mu \alpha}
\end{array}
\right.
\end{equation}

The formulae (\ref{35}) describe the action of $(\textbf{1},
\textbf{B}_\alpha, \textbf{B}_{\mu \alpha})$ on the space
$(\{\textbf{A}_\mu \}, \{\textbf{A}_{\mu \alpha}\})$. Our purpose
is to describe explicitly the projection
\begin{equation}\label{40}
\tilde{\pi} : j^1(T^*(M) \otimes \xi^* \otimes \xi)_0
\longrightarrow j^1(T^*(M) \otimes \xi^* \otimes
\xi))_0/Aut^V(\xi) \, .
\end{equation}
First of all $j^0(T^*(M) \otimes \xi^* \otimes \xi)_0$ is a
homogeneous space. So we can take $\textbf{B}_\mu = -
\textbf{A}_\mu$ and acting by $(\textbf{1}, -\textbf{A}_\mu,
\textbf{B}_{\mu \alpha})$ on $j^1(A_\mu)_0$ we obtain
$(\textbf{A}_\mu , \textbf{A}_{\mu \alpha}) \mapsto (0,
\tilde{\textbf{A}}_{\mu \alpha})$, where $\tilde{\textbf{A}}_{\mu
\alpha} = \textbf{A}_{\mu \alpha} - \textbf{A}_\mu
\textbf{A}_\alpha + \textbf{B}_{\mu \alpha}$. Due to the
homogeneity of $j^0(T^*(M) \otimes \xi^* \otimes \xi)_0$ there is
no obstruction at the level of the 0-jets and we go to the level
of 1-jets. So we consider only elements of the type  $(\textbf{1},
0, \textbf{B}_{\mu \alpha})$ acting on $(0,
\tilde{\textbf{A}}_{\mu \alpha})$, i.e. $\tilde{\textbf{A}}_{\mu
\alpha} \mapsto \tilde{\textbf{A}}_{\mu \alpha} + \textbf{B}_{\mu
\alpha}$.

Due to the symmetry of the matrices $\textbf{B}_{\mu \alpha}$ with
respect to $\mu, \alpha$ the factor space $(\{\textbf{A}_{\mu
\alpha}\})/(\{\textbf{B}_{\mu \alpha}\})$ is represented by the
space of matrices $\{\textbf{A}_{\mu \alpha}\}$ antisymmetric with
respect to $\mu, \alpha$ and the projection is the
antisymmetrization with respect to $\mu, \alpha$:
$\tilde{\textbf{A}}_{\mu \alpha} \mapsto
\tilde{\textbf{A}}_{\alpha \mu } - \tilde{\textbf{A}}_{\mu
\alpha}$. Finally we obtain
\[
(\textbf{A}_\mu , \textbf{A}_{\mu \alpha}) \mapsto (0,
\tilde{\textbf{A}}_{\mu \alpha}) \mapsto
\tilde{\textbf{A}}_{\alpha \mu } - \tilde{\textbf{A}}_{\mu \alpha}
\]
\begin{equation}\label{50}
\tilde{\textbf{A}}_{\alpha \mu } - \tilde{\textbf{A}}_{\mu \alpha}
= \partial_\mu \tilde{\textbf{A}}_\alpha - \partial_\alpha
\tilde{\textbf{A}}_\mu = \partial_\mu A_\alpha (0) -
\partial_\alpha A_\mu (0)+ [A_\mu (0), A_\alpha (0)] = F_{\mu \nu}^\nabla (0)\, .
\end{equation}
The differential operator $\pi$ corresponding to $\tilde{\pi}$
(\ref{40}) is the Yang-Mills curvature tensor,
\[
\pi(\nabla) = F^\nabla \,.
\]
For a flat connection the curvature tensor is equal to zero. For a
connection $\nabla$ the condition
\[
F^{\nabla} \equiv \tilde{\pi}(j^1(A)) \neq 0
\]
is an obstruction for $\nabla$ to be a flat connection.

\textbf{Example 2.5} - Almost complex structure on an even
dimensional
    manifold.

 Following the general scheme considered in Section 3 we will
 describe the Nijenhuis differential operator as an obstruction
 to the (local) equivalence of almost complex structures on
 the even dimensional manifold $M$. The nonvanishing Nijenhuis
 tensor for some almost complex structure is an obstruction to its
 integrability, i.e. to its equivalence to the canonical almost
 complex structure for which the Nijenhuis tensor vanishes.

 In this example the bundle under consideration is the vector bundle
 $T^*(M) \otimes T(M)$. An almost complex structure $J$ on $M$ is
 an element of $C^\infty(T^*(M) \otimes T(M))$ with the property
 $J^2 = -\mathbf{1}$. In coordinates $\{x^\mu\}$ centered at a point $x_0
 \in M$, $x^\mu(x_0) = 0$, the 1-jet of an almost complex
 structure $J$ reads
 \[
 j^1(J)_{0 \nu}^{\,\,\mu} (x) = \textbf{J}_\nu^{\,\mu} + \textbf{C}_{\nu \rho}^\mu x^\rho \, ,
 \]
where $\textbf{J}_\mu^\nu = J_\mu^\nu(0)$, $\textbf{C}_{\nu
\rho}^\mu = \partial_\rho J_\nu^\mu(0)$, $\textbf{J}_\alpha^\mu
\textbf{J}_\nu^\alpha = -\delta_\nu^\mu$, and
$\textbf{J}_\alpha^\mu \textbf{C}_{\nu \rho}^\alpha +
\textbf{C}_{\alpha \rho}^\mu \textbf{J}_\mu^\alpha = 0$.

The functional group is the group $Diff(M)$ acting by the tangent
lifting. The stationary group is the group $Diff(M)_{x_0}$ - the
group of the diffeomorphisms with a stable point $x_0$. For the
2-jet of a diffeomorphism $\varphi \in Diff(M)_{x_0}$ we have
\[
j^2(\varphi)_0^\mu(x) = \textbf{B}_\alpha^\mu x^\alpha +
\frac{1}{2} \textbf{B}_{\alpha \beta}^\mu x^\alpha x^\beta \, ,
\]
where $\textbf{B}_\alpha^\mu = \partial_\alpha \varphi^\mu(0)$,
$\textbf{B}_{\alpha \beta}^\mu = \partial_\alpha \partial_\beta
\varphi^\mu(0)$ and $\det\{\textbf{B}_\alpha^\mu \} \neq 0$.

Here again the crucial role play the diffeomorphisms $\varphi$
with $\textbf{B}_\alpha^\mu = \delta_\alpha^\mu$ i.e. with a
tangent lifting $\varphi^T: T_{x_0}(M) \longrightarrow T_{x_0}(M)$
equal to the identity map. We will consider only diffeomorphisms
of this kind. For the 2-jets we have
\[
j^2(\varphi_0)^\mu(x) = x^\mu + \frac{1}{2}\textbf{B}^\mu_{\alpha
\beta}x^\alpha x^\beta \,.
\]
The action of these diffeomorphisms on the 1-jet of the almost
complex structure $J$ at the point $0$ is given by
\begin{equation}\label{60}
\left|
\begin{array}{l} \textbf{J}_\nu^\mu \longrightarrow \textbf{J}_\nu^\mu \\
\textbf{C}_{\nu \rho}^\mu \longrightarrow \textbf{C}_{\nu
\rho}^\mu + \textbf{J}_\alpha^\mu \textbf{B}_{\nu \rho}^\alpha -
\textbf{B}_{\alpha \rho}^\mu \textbf{J}_\nu^\alpha \,.
\end{array}
\right.
\end{equation}
Due to (\ref{60}) we consider only the space $W$ of 1-jets of the
almost complex structure $J$ at the point $0$ over a fixed
$\{\textbf{J}_\nu^\mu\} \equiv \textbf{J}$. This space is
parameterized by $\{\textbf{C}_{\nu \rho}^\mu\}$. The description
of the space $W$ and of the factor-space of $W$ with respect to
the action (\ref{60}) is more visual if we consider
$\{\textbf{C}_{\nu \rho}^\mu\}$ and $\{\textbf{B}_{\alpha
\beta}^\mu\}$ as elements of the space $L^* \otimes L^* \otimes
L$, where $L$ is a vector space with $\dim(L) = m$, or
equivalently as bilinear forms: $L\times L \longrightarrow L$.

In this interpretation
\[
W=\{\textbf{C} \in L^* \times L^* \times L \mid
\textbf{J}\textbf{C}(u, v) + \textbf{C}(\textbf{J}u, v) = 0, u, v
\in L\}\, .
\]
Let us define a map $K: S^2L^* \otimes L \longrightarrow W$ by
\begin{equation}\label{70}
K(\textbf{B})(u, v) = \textbf{JB}(u, v) - \textbf{B}(\textbf{J}u,
v)\, .
\end{equation}
In theses notations the action (\ref{60}) of the diffeomorphisms
on the space $W$ reads
\[
W \ni \textbf{C} \longrightarrow \textbf{C} + K(\textbf{B})\,,
\,\textbf{B} \in S^2L^* \otimes L\,.
\]

Our purpose is to describe the projection
\[
\pi: W \longrightarrow W/K(S^2L^* \otimes L)
\]
By $s: W \longrightarrow S^2L^* \otimes L$ we denote the
symmetrization
\[
s(\textbf{C})(u, v) = \frac{1}{2}(\textbf{C}(u, v)+ \textbf{C}(v,
u)) \, .
\]
The map $A \equiv s \circ K: S^2L^* \otimes L \longrightarrow
S^2L^* \otimes L$ is an invertible map but not equal to the
identity. The space $W$ splits into the following direct sum
\begin{equation}\label{80}
W = K(S^2L^* \otimes L) \oplus ker(s) \, .
\end{equation}
The projections on the first and on the second term in (\ref{80})
are the following
\[
K \circ A^{-1} \circ s: W \longrightarrow K(S^2L^* \otimes L)
\]
and
\[
\textbf{1} - K \circ A^{-1} \circ s :W \longrightarrow ker(s) \, .
\]
It is easy to calculate that
\[
(K \circ A^{-1} \circ s)(\textbf{C}) (u, v) =
\frac{1}{2}(\textbf{C}(u, v) +\textbf{C}(v, u) - \textbf{JC}(u,
\textbf{J}v) + \textbf{JC}(v, \textbf{J}u))
\]
and
\[
\begin{array}{ll}
(\textbf{1} -K \circ A^{-1}& \circ s)(\textbf{C}) (u, v) = \\
&\frac{1}{2}(\textbf{C}(u, v) -\textbf{C}(v, u) - \textbf{JC}(u,
\textbf{J}v) + \textbf{JC}(v, \textbf{J}u)) =
-\frac{1}{2}\textbf{J}.\textbf{N}(J) \, .
\end{array}
\]
The projection
\[
-\frac{1}{2} \textbf{J}.N(\textbf{J}): W \longrightarrow ker(s)
\approx W/K(S^2L^* \otimes L)
\]
is the projection we are looking for. The operator $J(x)$ is an
invertible operator so the important information is carried by
$N(J)$. For the canonical almost complex structure $N(J) = 0$.
That's why $N(J) \neq 0$ is an obstruction to the (local)
equivalence of the almost complex structure $J$ to the canonical
one, i.e. to its integrability.

\section{Definition of the Curvature of a Superconnection as an Obstruction}\label{sec:5}

Let $\nabla_s = \nabla + \chi$ be a superconnection \cite{Quillen}
on a $\mathbb{Z}_2$-graded bundle $\xi$. Let $\{x^\mu\}$ be
coordinates on the base $M$ and $\{e_{+a}\}$, $\{e_{-i}\}$ be a
basis in $\xi$, compatible with the $\mathbb{Z}_2$-grading. In
coordinates
\[
\nabla_\mu = \partial_\mu + A_\mu \, ; \, A_\mu = A_{+\mu} +
A_{-\mu}
\]
\[
A_+ = A_{+\mu a}^{\, \,\, \,b} \, \d x^\mu \otimes e_+^a \otimes
e_{+b} \, , \, A_- = A_{- \mu i}^{\, \,\, \,j}\, \d x^\mu \otimes
e^{\, \,i}_- \otimes e_{-j}\, ,
\]
\begin{equation}\label{90}
\chi = \chi_{+-i}^{\,\, \,\, \, \,\, \,a}\, e_-^{\, \,i} \otimes
e_{+\,a} + \chi_{-+a}^{\, \, \,\,\, \,\, \,i}\, e_+^{\, \,a}
\otimes e_{-i}\, ,
\end{equation}
\[
(\chi_{+-} + \chi_{-+} , A_+ + A_-) \in \Omega^{0, 1} (\xi^*
\otimes \xi)_-\, .
\]
If the origin $\nabla_{s0\,\,\mu} = \partial_\mu + 0$ in the space
of the superconnections is fixed the elements of
$\Omega^{0,1}(\xi^* \otimes \xi)_-$ are in one-to-one
correspondence with the superconnections. So in the case of
superconnections the bundle under consideration is $(\xi^* \otimes
\xi)_- \oplus (T^*(M) \otimes \xi^* \otimes \xi)_-$. The 1-jet at
the point $0$ of a superconnection, or more precisely, of its
components reads
\[
\begin{array}{llll}
j^1(\chi_{+-})_0(x) = & \mbox{\boldmath $\chi$}_{+-}+
\mbox{\boldmath $\chi$}_{+-\mu}x^\mu ; \, & \mbox{\boldmath
$\chi$}_{+-} = \chi_{+-}(0)  , \, & \mbox{\boldmath $\chi$}_{+-
\mu} =
\partial_\mu \chi_{+-}(0) \\
j^1(\chi_{-+})_0(x) = & \mbox{\boldmath
$\chi$}_{-+}+\mbox{\boldmath $\chi$}_{-+\mu}x^\mu  ; \, &
\mbox{\boldmath $\chi$}_{-+} = \chi_{-+}(0)  , \, &
\mbox{\boldmath $\chi$}_{-+\mu} =
\partial_\mu \chi_{-+}(0)\\
j^1(A_{+ \mu})_0(x) =  & \textbf{A}_{+ \mu} + \textbf{A}_{+ \mu
\rho}x^\rho ; \, & \textbf{A}_{+ \mu} = A_{+ \mu}(0) , \,
&\textbf{A}_{+ \mu \rho} = \partial_\rho A_{+ \mu}(0) \\
j^1(A_{- \mu})_0(x) =  & \textbf{A}_{- \mu} + \textbf{A}_{- \mu
\rho}x^\rho ; & \, \textbf{A}_{- \mu} = A_{- \mu}(0) , \,
&\textbf{A}_{- \mu \rho} = \partial_\rho A_{- \mu}(0) \\
\end{array}
\]
The set $(\mbox{\boldmath $\chi$}_{+-}, \mbox{\boldmath
$\chi$}_{+- \rho}, \mbox{\boldmath $\chi$}_{-+} , \mbox{\boldmath
$\chi$}_{-+ \rho}, \textbf{A}_{+ \mu}, \textbf{A}_{+ \mu \rho},
\textbf{A}_{- \mu}, \textbf{A}_{- \mu \rho})$ parameterizes the
space $j^1((\xi^* \otimes \xi)_- \oplus (T^*(M) \otimes \xi^*
\otimes \xi)_-)_0$.

The "functional" group is $Aut^V(\xi_+ \oplus \xi_-)$. For every
element $ \varphi \in Aut^V(\xi_+ \oplus \xi_-)$ we have $ \varphi
= \varphi_+ + \varphi_-$. The action of $\varphi$ on $\Omega^{0,
1}(\xi^* \otimes \xi)$ is given by
\begin{equation}\label{93}
\begin{array}{ll}
(\chi_{+-}&+ \chi_{-+} , A_+ + A_-) \longrightarrow
\varphi^*(\chi_{+-} + \chi_{-+} , A_+ + A_-) = (\varphi_+^{-1}.
\chi_{+-} . \varphi_- + \\
&+\varphi_-^{-1} . \chi_{-+} . \varphi_+ , \varphi_+^{-1} . A_{+
\mu} . \varphi_+ + \varphi_+^{-1}.
\partial_\mu \varphi_+ ,  \varphi_-^{-1} . A_{- \mu} . \varphi_- +
\varphi_-^{-1} .
\partial_\mu \varphi_-)
\end{array}
\end{equation}
The 2-jet of $\varphi \in Aut^V(\xi_+ \oplus \xi_-)$ at the point
$0$ reads
\[
\begin{array}{lll}
j^2(\varphi_+)(x) = \mathbf{1} + \mbox{\boldmath $\varphi$}_{+
\rho }x^\rho + \frac{1}{2} \mbox{\boldmath $\varphi$}_{+ \rho
\sigma} x^\rho x^\sigma ; \, & \mbox{\boldmath $\varphi$}_{+ \rho}
=
\partial _\rho \varphi_+ (0) , \, & \mbox{\boldmath $\varphi$}_{+
\rho \sigma} = \partial_\rho \partial_\sigma \varphi_+(0) \\

j^2(\varphi_-)(x) = \mathbf{1} + \mbox{\boldmath $\varphi$}_{-
\rho }x^\rho + \frac{1}{2} \mbox{\boldmath $\varphi$}_{- \rho
\sigma} x^\rho x^\sigma ; \, & \mbox{\boldmath $\varphi$}_{- \rho}
=
\partial _\rho \varphi_- (0) , \, & \mbox{\boldmath $\varphi$}_{-
\rho \sigma} = \partial_\rho \partial_\sigma \varphi_-(0) \\
\end{array}
\]
As in the previous examples (see Section 4) we consider only
automorphisms with 2-jets beginning with the identity operator.
From (\ref{90}) and (\ref{93}) for the action of an automorphism
$\varphi$ on the 1-jet of the superconnection we obtain
\[
\left|
\begin{array}{lll}
\mbox{\boldmath $\chi$}_{+-} & \longrightarrow  & \mbox{\boldmath $\chi$}_{+-} \\
\mbox{\boldmath $\chi$}_{+- \mu} & \longrightarrow &
\mbox{\boldmath $\chi$}_{+- \mu} - \mbox{\boldmath $\varphi$}_{+
\mu}. \mbox{\boldmath $\chi$}_{+-} + \mbox{\boldmath $\chi$}_{+-}.
\mbox{\boldmath $\varphi$}_{- \mu} \\
\end{array}
\right.
\]

\begin{equation}\label{95}
\left|
\begin{array}{lll}
\mbox{\boldmath $\chi$}_{-+} & \longrightarrow  & \mbox{\boldmath $\chi$}_{-+} \\
\mbox{\boldmath $\chi$}_{-+ \mu} & \longrightarrow &
\mbox{\boldmath $\chi$}_{-+ \mu} - \mbox{\boldmath $\varphi$}_{-
\mu} .\mbox{\boldmath $\chi$}_{-+} + \mbox{\boldmath $\chi$}_{-+}.
\mbox{\boldmath $\varphi$}_{+ \mu} \\
\end{array}
\right. \end{equation}

\[
\left|
\begin{array}{lll}
\mathbf{A}_{+\mu} & \longrightarrow  & \mathbf{A}_{+\mu} +
\mbox{\boldmath $\varphi$}_{+ \mu} \\
\mathbf{A}_{+ \mu \rho} & \longrightarrow & \mathbf{A}_{+ \mu
\rho} - \mbox{\boldmath $\varphi$}_{+ \rho}. \mathbf{A}_{+ \mu} +
\mathbf{A}_{+ \mu}.\mbox{\boldmath $\varphi$}_{+ \rho}-
\mbox{\boldmath $\varphi$}_{+ \rho}. \mbox{\boldmath $\varphi$}_{+
\mu} +
\mbox{\boldmath $\varphi$}_{+\mu \rho} \\
\end{array}
\right.
\]

\[
\left|
\begin{array}{lll}
\mathbf{A}_{-\mu} & \longrightarrow  & \mathbf{A}_{-\mu} +
\mbox{\boldmath $\varphi$}_{- \mu} \\
\mathbf{A}_{- \mu \rho} & \longrightarrow & \mathbf{A}_{- \mu
\rho} - \mbox{\boldmath $\varphi$}_{- \rho} .\mathbf{A}_{- \mu} +
\mathbf{A}_{- \mu}.\mbox{\boldmath $\varphi$}_{- \rho}-
\mbox{\boldmath $\varphi$}_{- \rho}. \mbox{\boldmath $\varphi$}_{-
\mu} +
\mbox{\boldmath $\varphi$}_{-\mu \rho} \\
\end{array}
\right.
\]

We choose $\mbox{\boldmath $\varphi$}_{+\mu} =
-\textbf{A}_{+\mu}$, $\mbox{\boldmath $\varphi$}_{-\mu} =
-\textbf{A}_{-\mu}$ and the upper transformations lead to
\begin{equation}\label{97}
\begin{array}{ll}
(\mbox{\boldmath $\chi$}_{+-}, \mbox{\boldmath $\chi$}_{+- \mu},
&\mbox{\boldmath $\chi$}_{-+} , \mbox{\boldmath $\chi$}_{-+ \mu},
\textbf{A}_{+ \mu}, \textbf{A}_{+ \mu \rho}, \textbf{A}_{- \mu},
\textbf{A}_{- \mu \rho}) \longrightarrow
\\ & (\mbox{\boldmath $\chi$}_{+-}, \tilde{\mbox{\boldmath $\chi$}}_{+- \mu}, \mbox{\boldmath $\chi$}_{-+} ,
\tilde{\mbox{\boldmath $\chi$}}_{-+ \mu}, 0, \tilde{\textbf{A}}_{+
\mu \rho}, 0, \tilde{\textbf{A}}_{- \mu \rho})
\end{array}
\end{equation}

\[
\left|
\begin{array}{ll}
\tilde{\mbox{\boldmath $\chi$}}_{+-\mu} =  & \mbox{\boldmath
$\chi$}_{+-\mu} +\textbf{A}_{+\mu}. \mbox{\boldmath $\chi$}_{+-} -
\mbox{\boldmath $\chi$}_{+-} .\textbf{A}_{-\mu} \\
\tilde{\mbox{\boldmath $\chi$}}_{-+\mu} =  & \mbox{\boldmath
$\chi$}_{-+\mu} +\textbf{A}_{-\mu} .\mbox{\boldmath $\chi$}_{-+} -
\mbox{\boldmath $\chi$}_{-+} .\textbf{A}_{+\mu} \\
\tilde{\textbf{A}}_{+ \mu \rho} = & \textbf{A}_{+ \mu \rho}- \textbf{A}_{+\mu} .\textbf{A}_{+\rho} \\
\tilde{\textbf{A}}_{- \mu \rho} = & \textbf{A}_{- \mu \rho}+ \textbf{A}_{-\mu} .\textbf{A}_{-\rho} \\
\end{array}
\right.
\]

Next we consider automorphisms which acting according (\ref{95})
preserve the special form (\ref{97}) of the 1-jet of the
superconnection. The 1-jets of these automorphisms have the form
$(\mathbf{1}, 0, \mbox{\boldmath $\varphi$}_{+ \mu \rho};
\mathbf{1} , 0 , \mbox{\boldmath $\varphi$}_{- \mu \rho})$. Their
action on
\[
(\mbox{\boldmath $\chi$}_{+-}, \tilde{\mbox{\boldmath $\chi$}}_{+-
\rho}, \mbox{\boldmath $\chi$}_{-+} , \tilde{\mbox{\boldmath
$\chi$}}_{-+ \rho}, 0, \tilde{\textbf{A}}_{+ \mu \rho}, 0,
\tilde{\textbf{A}}_{- \mu \rho})
\]
is given by
\[
\begin{array}{ll}
(\mbox{\boldmath $\chi$}_{+-}, & \tilde{\mbox{\boldmath
$\chi$}}_{+- \rho}, \mbox{\boldmath $\chi$}_{-+} ,
\tilde{\mbox{\boldmath $\chi$}}_{-+ \rho}, 0, \tilde{A}_{+ \mu
\rho}, 0, \tilde{A}_{- \mu \rho})
\longrightarrow \\
& (\mbox{\boldmath $\chi$}_{+-}, \tilde{\mbox{\boldmath
$\chi$}}_{+- \rho}, \mbox{\boldmath $\chi$}_{-+} ,
\tilde{\mbox{\boldmath $\chi$}}_{-+ \rho}, 0,
\tilde{\textbf{A}}_{+ \mu \rho}+ \mbox{\boldmath $\varphi$}_{+ \mu
\rho}, 0, \tilde{\textbf{A}}_{- \mu \rho}+ \mbox{\boldmath
$\varphi$}_{- \mu \rho}).
\end{array}
\]
Due to the symmetry of the matrices $\mbox{\boldmath
$\varphi$}_{\pm \mu \rho}$ with respect to $\mu$, $\rho$ the
factor space, we are looking for, is obtained by
antisymmetrization of $\{ \tilde{\textbf{A}}_{\pm \mu \rho}\}$.
The canonical projection
\[
\begin{array}{ll}
\tilde{\pi} : j^1((\xi^* \otimes & \xi)_- \oplus (T^*(M) \otimes
\xi^*
\otimes \xi)_-) \longrightarrow \\
& j^1((\xi^* \otimes \xi)_- \oplus (T^*(M) \otimes \xi^* \otimes
\xi)_-)/ Aut^V(\xi_+ \oplus \xi_-)
\end{array}
\]
is the composition of (\ref{97}) and the antisymmetrization of
$\{\tilde{\textbf{A}}_{\pm \mu \rho} \}$ with respect to $\mu ,
\rho$,
\[
\begin{array}{ll}
(\mbox{\boldmath $\chi$}_{+-}, &\mbox{\boldmath $\chi$}_{+- \mu},
\mbox{\boldmath $\chi$}_{-+} , \mbox{\boldmath $\chi$}_{-+ \mu},
\textbf{A}_{+ \mu}, \textbf{A}_{+ \mu \rho}, \textbf{A}_{- \mu},
\textbf{A}_{- \mu \rho}) \longrightarrow
\\ & (\mbox{\boldmath $\chi$}_{+-},
\tilde{\mbox{\boldmath $\chi$}}_{+- \mu}, \mbox{\boldmath
$\chi$}_{-+} , \tilde{\mbox{\boldmath $\chi$}}_{-+ \mu}, 0,
\tilde{\textbf{A}}_{+ \mu \rho}, 0, \tilde{\textbf{A}}_{-
\mu \rho}) \longrightarrow \\
& (\mbox{\boldmath $\chi$}_{+-}, \tilde{\mbox{\boldmath
$\chi$}}_{+- \mu}, \mbox{\boldmath $\chi$}_{-+} ,
\tilde{\mbox{\boldmath $\chi$}}_{-+ \mu}, 0, \tilde{\textbf{A}}_{+
\rho \mu }-\tilde{\textbf{A}}_{+ \mu \rho} , 0,
\tilde{\textbf{A}}_{- \rho \mu }-\tilde{\textbf{A}}_{- \mu
\rho}) = \\
&(\chi_{+-}(0), \nabla_\mu \chi_{+-}(0), \chi_{-+}(0), \nabla_\mu
\chi_{-+}(0),
\partial_\mu A_{+ \rho}(0) - \partial_\rho A_{+ \mu}(0) + \\
&[A_{+ \mu}(0), A_{+ \rho}(0)],  \partial_\mu A_{- \rho}(0) -
\partial_\rho A_{- \mu}(0) + [A_{- \mu}(0), A_{- \rho}(0)]).
\end{array}
\]
In short notations
\[
\tilde{\pi}: j^1(\chi , \nabla) \longrightarrow (\chi, \nabla \chi
, F^\nabla) \in \Omega^{0, 1, 2}(\xi^* \otimes \xi)\, .
\]
Finally, we claim that the natural notation of curvature of a
superconnection $\nabla_s = \nabla + \chi$ on a
$\mathbb{Z}_2$-graded vector bundle $\xi$ is the operator
\[
\pi(\nabla_s) = (\chi, \nabla(\chi), F^\nabla) \in \Omega^{0, 1
,2}(\xi^* \otimes \xi) \,.
\]
This notion differs from the expression $(\chi^2, \nabla(\chi),
F^\nabla)$ obtained by purely algebraic analogy with the curvature
of a linear connection.

Our notion of a curvature of a superconnection may be of interest
to the models of interacting particles in the supersymmetrical
field theories where one of the expressions in the Lagrangian is
the square of the supercurvature.

The obstructions we have considered are related to the action of
"functional" groups on $k$-jets of smooth sections of some vector
bundles. The study of the orbits of this action is usually called
a study of the singularities of smooth maps or "Catastrophe
Theory" in the terminology of Ren$\mathrm{\acute{e}}$ Thom (see
\cite{Arnold}, \cite{Brocker}). The title of our paper is in the
Ren$\mathrm{\acute{e}}$ Thom's terminology.\\ \\
{\bf In memoriam of our dear friend and colleague \\
{\large Ventzeslav Rizov}}

\end{document}